\documentclass[a4paper,10pt]{article}
\usepackage[utf8]{inputenc}

\usepackage{xcolor}

\title{ Quantization of Yang-Mills Theory in de Sitter Spacetime }
\author{ Aasiya Shaikh$^1$,    Mir Faizal$^{2,3}$, Naveed Ahmad Shah$^{3,4}$ \\\\
\small$^1$Raja Ramanna Centre for Advanced Technology,  Indore-452013, Madhya Pradesh, India.\\\small
\small $^2$Irving K. Barber School of Arts and Sciences, University of British Columbia - Okanagan, 
\\\small Kelowna, BC V1V 1V7, Canada.\\\
\small$^{3}$Canadian Quantum Research Center 204-3002  32 Ave Vernon, BC V1T 2L7 Canada. 
\\\
\small$^{4}$ Department of Physics, Jamia Millia Islamia, New Delhi - 110025, India}
\date{}
\begin{document}
\maketitle 
\begin{abstract}
In this paper, we analyze the quantization of Yang-Mills theory in the de Sitter spacetime. 
It is observed that the  Faddeev--Popov ghost propagator is divergent in this spacetime. However, this divergence is removed by using an effective propagator, which is suitable for perturbation theory. To show that the quantization of
Yang--Mills theory in the de Sitter is consistent, we quantize it using first-class constraints in the temporal gauge.  We also demonstrate that this is equivalent to quantizing the theory in the Lorentz gauge.  
 \end{abstract}

\section{Introduction}
It is expected that our universe will asymptotically approach de Sitter spacetime, due to the observations from type I supernovae
\cite{super, super2, super0, super4, super5}. At such large scales, the quantum effects can be neglected, and the system can be described using classical cosmology. 
However, quantum field theory in de Sitter spacetime is important as the universe in inflationary cosmology is also represented by de Sitter spacetime  \cite{cos, cos0, cos1, cos2}. 
Thus, various interesting models of inflation have been studied using de Sitter spacetime. It is used  
in brane-antibrane models \cite{z, za} and $D3/D7$ systems \cite{z2, z2a}, where  the inflaton field corresponds to an open string. 
The de Sitter spacetime is also important in Kahler moduli \cite{z4, z4a} and fibre inflation \cite{z7}, where inflaton fields correspond to 
closed strings. In these models the realization of inflation depends crucially 
on the uplifting mechanism for de Sitter moduli stabilization \cite{coss}.  
So, quantum field theory in de Sitter spacetime is both interesting and important. 

It may be noted that Yang-Mills theory has also been used to model inflationary cosmology \cite{yang}. In this analysis, a ten-dimensional
 Einstein-Yang-Mills theory has been used to study cosmological solutions. It has been demonstrated that in such a theory, it is possible 
 to obtain cosmological solutions with static extra dimensions. Furthermore, for such cosmological models, the scale factor of the four-dimensional Friedmann-Lemaitre-Robertson-Walker metric is an exponential function of time.
   Thus, it has been argued that such a model can be used for analyzing inflation, and so it is important to study Yang-Mills theories in de Sitter spacetime. 
 The Yang-Mills theory in de Sitter spacetime is also important for analyzing physical systems using the gauge/gravity duality. 
 In the usual gauge/gravity, the quantum field theory in AdS spacetime is dual to a CFT in flat spacetime \cite{adscft}. 
 It is important to analyze the cosmological singularities using this duality, but it has been observed that the boundary theory dual to such 
 singularities also becomes singular \cite{s1, s2}. However, it is possible to analyze such singularities, if the dual field theory is taken 
 as a gauge theory on de Sitter spacetime \cite{s4, s5}. This makes it important to analyze the Yang-Mills theory in de Sitter spacetime.

Even though it is important to study quantum field theory in de Sitter spacetime, there are several issues with the 
consistency of the quantum field theory in this spacetime. There is a problem with the 
linearization instabilities in de Sitter spacetime \cite{2}. 
There are also problems with 
certain values of gauge parameters for perturbative quantum gravity \cite{w2,w1}. 
There are several other problems relating to infrared divergence, 
and it has been argued that it would not be possible to integrate by parts due to such infrared divergences  \cite{1, q1, q2, p}.
Even though there are problems with perturbative quantum gravity, it is possible to analyze the perturbative quantum 
gravity in the first order formalism as a gauge theory of spin connection \cite{grav}.
In fact, for higher curvature terms, this theory resembles the Yang-Mills theory, with the gauge group being the Lorentz group. 
The inflation has also been studied using gravity with higher order curvature terms \cite{curv}. 
Thus, such a system can also be analyzed as a Yang-Mills theory, and consistent results can be obtained using such a system. 

However, there are certain issues even with the Yang-Mills theories in de Sitter spacetime. 
When Yang-Mills theory is quantized using the path integral approach ghost fields are introduced into the theory. It has been shown \cite{Fai1}
that the ghost propagators are infrared divergent in de Sitter spacetime. This issue can be resolved by introducing a mass that is taken to zero at 
the end of the perturbative calculations. This mass term breaks BRST invariance but it has been shown \cite{Gibb1} that the zero modes, which cause the 
infrared divergences can be removed in a BRST invariant way, producing a theory equivalent to the one obtained by adding a mass term. 
However, integration by parts was used to obtain this result, but it has also been argued that it might not be possible 
to use integration by parts in perturbative calculations in de Sitter space  \cite{1, q1, q2, p}. So, it is important to understand if Yang-Mills theories can 
be consistently quantized in de Sitter spacetime. This motivates us to use constraint quantization \cite{Dir1, Dir2} to quantize Yang-Mills theory in de Sitter spacetime. 
We will be able to demonstrate that the theory can be consistently quantized using constraint quantization. It is already known that if a theory can be quantized using constraint quantization, then it is consistent to quantize that theory using the BRST quantization, as both these approaches have been demonstrated to be physically equivalent to each other \cite{g1}. Thus, if the Yang-Mills theory is consistently quantized using constraint quantization, then it is naturally consistent to quantize it using the BRST formalism. It may be noted that the BRST quantization of Yang-Mills theory in de Sitter spacetime has already been studied \cite{Gibb1}, and here we show that it is consistent to use such an approach by quantizing it using constraint quantization. We will first show that the theory can be consistently quantized in the temporal gauge, and then show that this is equivalent to quantizing it in the Lorentz gauge.  As the removal of the infrared divergence has been studied in the Lorentz gauge \cite{Gibb1}, we can argue that it is consistent to use such a formalism.

\section{Effective Propagator} 
The de Sitter spacetime has the topology $S^N \times R$. So, it is possible to take a  
$N$ dimensional space-like slice $\Sigma = S^N$ though the spacetime. It may be noted that for analyzing Yang-Mills theory
using constraint analysis, it would not be required to restrict the space-like slice to $S^N$, and the analysis will hold 
for any spacetime with the topology $\Sigma \times R$. 
So, the metric for such a spacetime can be written as 
\begin{equation}
ds^2= g_{\mu\nu} dx^\mu dx^\nu = -dt^2+h_{ij}(t,x)dx^idx^j\;.
\end{equation}
The  action  for the Yang-Mills theory is given by
\begin{equation}
S = \int d^{d+1} x \sqrt{h} L=-\int d^{d+1} x  \frac{1}{4}\sqrt{h}F^{\mu\nu}_AF_{\mu\nu}^A\;,
\label{Lag}
\end{equation}
where
\begin{equation}
F_{\mu\nu}^A=\partial_{\mu}A_{\nu}^A-\partial_{\nu}A_{\mu}^A+gC^A_{\;\; BC}A^B_{\mu}A^C_{\nu}\;,
\end{equation}
$C^A_{\;\; BC}$ are the structure constants of the group defined by
\begin{equation}
[t_A,t_B]=iC^C_{\;\; AB}t_C\;,
\end{equation}
$t_A$ are the generators of the group and $g$ is the coupling constant.

This theory can be quantized using the path integral by first choosing a gauge, for example, the Lorentz gauge $\nabla^\mu A_\mu^A =0$. 
This gauge fixing condition can be implemented at a quantum level by adding a gauge fixing term and a ghost term to the original action. 
Thus, the total effective action for Yang-Mills theory can be written as 
\begin{eqnarray}
S_{gf} + S_{gh} = \int d^{d+1} x \sqrt{h} \left[B_A \nabla^\mu A^{\mu} + \frac{\alpha}{2} B^A B_A + \nabla^\mu \bar c_A D_\mu c^A \right], 
\end{eqnarray}
where $c^A$ and $\bar c^A$ is the ghost and anti-ghost fields. The gauge conditions are implemented using the auxiliary field $B_A$. 
Now using the total action, which is given by the sum of the original action, the gauge fixing term, and the ghost term, 
$S_T = S + S_{gf} + S_{gh}$, the path integral 
can be defined as
\begin{eqnarray}
 Z = \int D A \ Dc \ D\bar c \ D B e^{i S_T}. 
\end{eqnarray}
The correlation function can be calculated from this path integral using the usual methods.
The propagator for the gauge fields can be directly calculated in de Sitter spacetime. 
However, in de Sitter spacetime, 
there is a problem in quantizing the theory in this way, as it has been demonstrated that in four dimensions the ghost propagator is
infrared divergent  \cite{Fai1}. Here we will generalize this result to  $d+1$ dimensions, and we will demonstrate that the ghost propagator 
is infrared divergent in $d+1$ dimensions. However, it is possible to obtain an effective propagator in $d+1$ dimensions by 
adding a mass term and then setting the mass term to zero at the end of the calculations. 
Thus, we first write the  equation of motion for the ghost and anti-ghost fields, 
\begin{eqnarray}
 \nabla^\mu \nabla_\mu c^A =0, && \nabla^\mu \nabla_\mu \bar c^A =0.  
\end{eqnarray}
Now we can Euclideanize this system from $S^N \times R$ to $S^{N+1}$, and choose a Euclidean vacuum state for this theory. Then it is possible to write 
\begin{eqnarray}
 \langle 0| T [c^A (x) \bar c^B (x') ] |0 \rangle = i \delta^{AB} D_0 (x, x'), 
\end{eqnarray}
where $D_0 (x, x')$ would satisfy 
\begin{eqnarray}
 \nabla^\mu \nabla_\mu D_0 (x, x') = - \delta (x, x'). 
\end{eqnarray}
Now for  the spherical harmonics in $S^{N+1}$, we have 
\begin{eqnarray}
-\nabla^\mu \nabla_\mu Y^{L\sigma} = L (L+N) Y^{L\sigma}, 
\end{eqnarray}
here $\sigma$ denotes  all the labels other than $L$. 
We can write the $\delta (x, x')$ and $D_0 (x, x')$ on $S^{N+1}$ as 
\begin{eqnarray}
 \delta(x, x') &=& \sum_{L = 0}^\infty \sum_\sigma Y^{L\sigma} (x) Y^{L\sigma} (x'), \nonumber \\
 D_0 (x, x') &=& \sum_{L = 0}^\infty \sum_\sigma  k_L Y^{L\sigma} (x) Y^{L\sigma} (x'), 
\end{eqnarray}
where $k_L$ is a constant, and it is equal to $k_L = L (L +N)$. However, for $L =0$, this propagator is not well defined.
So, we regulate this propagator by adding a small mass $m^2$, and obtain 
\begin{eqnarray}
 D_{m^2}(x, x') &=&  \sum_{L = 0}^\infty \sum_\sigma \frac{Y^{L\sigma} (x) Y^{L\sigma} (x')}{L(L +N) + m^2 } \nonumber \\ 
 &=& \frac{1}{\sqrt{V} m^2} +  \sum_{L = 1}^\infty \sum_\sigma \frac{Y^{L\sigma} (x) Y^{L\sigma} (x')}{L(L +N) + m^2 }, 
\end{eqnarray}
where $V$ is the volume of a $S^{N+1}$. 
This propagator diverges in the zero mass limit, and the divergence comes from a constant mode. 
As these fields couple to the gauge fields through a derivative coupling, 
$   - i g C_{AB}^C \nabla^\mu \bar c^C A^A_\mu c^B$,   the constant  modes do not contribute to the perturbative calculations.
This is similar to the four-dimensional case \cite{Fai1}. Thus, we can write an effective propagator by subtracting the constant mode, and then 
taking the zero mass limit, 
\begin{eqnarray}
 D^{eff}_0 (x, x') = \lim_{m^2 \to 0}\left[ D_{m^2} (x, x') - \frac{1}{\sqrt{V} m^2}\right]. 
\end{eqnarray}
This propagator is finite in the zero mass limit and can be used to perform perturbative calculations. However, the zero mode can appear again 
in the full theory due to the BRST symmetry. It might be possible to generalize the analysis done to remove such modes in a BRST invariant way  \cite{Gibb1}, 
but this analysis depends on integration by parts. It has been argued that there are problems with performing integration by parts in de Sitter 
spacetime   \cite{1, q1, q2, p}. So, it is important to understand if it is possible to quantize the Yang-Mills theory in a consistent way in de Sitter spacetime. 
This can be done by quantizing the Yang-Mills theory in de Sitter spacetime using an alternative approach. So, in this paper, we will 
quantize the Yang-Mills theory in de Sitter spacetime using constraint quantization.

\section{Constraints}
In this section, we will again analyze the Yang-Mills theory on a spacetime with topology $S^N \times R$.  Here, we will specifically discuss the constraints in 
 the temporal gauge, and demonstrate that it is possible to consistently quantize a theory with topology $S^N \times R$ in 
 the temporal gauge. Now de Sitter spacetime has this topology, and it is this specific topology that causes the problems with the ghost propagators and the perturbative calculations (without removing the zero point mode) in de Sitter spacetime. So,  this analysis will demonstrate that it is possible to quantize Yang-Mills theory in de Sitter spacetime. 
So, first, we will review this theory as a classical field theory, and analyze its constraints. 
The canonical momenta $\Pi^{\mu}_A=\partial L/\partial \dot{A}_{\mu}^A$ are given by
\begin{equation}
\Pi^{\mu}_A=\sqrt{h}F_A^{\mu t}\;.
\label{Mom}
\end{equation}
There is therefore a set of primary constraints $\phi_A$ given by
\begin{equation}
\phi_A=\Pi_A^t\approx 0\;,
\end{equation}
where $\approx$ denotes a weak equality that can be imposed only after the Poisson brackets have been evaluated.
The canonical Hamiltonian is given by
\begin{equation}
H_c=\int d^dx\left[\frac{\Pi^k_A\Pi^A_k}{2\sqrt{h}}+\frac{1}{4}\sqrt{h}F^{ij}_AF^A_{ij}+gC^A_{\;\; BC}A^B_k\Pi^k_AA^C_t
-A_t^A\partial_k\Pi^k_A\right]\;.
\end{equation}
Consistency of the primary constraints requires that $\dot{\phi_A}=\{\phi_A,H_T\}\approx 0$, where $\{\;\;,\;\;\}$ denotes the Poisson bracket,
$H_T$ is the total Hamiltonian defined by $H_T=H_C+u^A\phi_A$ and $u^A$ are arbitrary parameters.
This gives the set of secondary constraints
\begin{equation}
\chi_A=D_k\Pi_A^k=\partial_k\Pi_A^k-gC^B_{\;\; CA}A^C_k\Pi^k_B\approx 0\;,
\end{equation}
where $D_k\Pi_A^k$ is the gauge covariant derivative of $\Pi^k_A$.
It can be shown that $\dot{\chi}_A=gC^B_{\;\; CA}A^C_t\chi_B\approx 0$, so that there are no more constraints. The constraints $\phi_A$ and
$\chi_A$ are first class constraints since $\{\phi_A(x),\phi_B(y)\}=0$, $\{\phi_A(x),\chi_B(y)\}=0$ and $\{\chi_A(x),\chi_B(y)\}
=gC^C_{\;\;AB}\chi_C(x)\delta(x,y)\approx 0$.
The canonical Hamiltonian can now be written as
\begin{equation}
H_c=\int d^dx\left[\frac{\Pi^k_A\Pi^A_k}{2\sqrt{h}}+\frac{1}{4}\sqrt{h}F^{ij}_AF^A_{ij}-A_t^A\chi_A\right]\;.
\end{equation}

In this section, the classical Yang-Mills theory will be quantized using Dirac's method for quantizing theories with first-class constraints. 
In this approach, the classical dynamical variables are promoted to operators that satisfy the standard commutation relations
\begin{equation}
[A_{\mu}^A(x),\Pi^{\nu}_B(y)]=i\delta^{\nu}_{\mu}\delta^A_B\delta^3(x,y)
\end{equation}
and
\begin{equation}
[A_{\mu}^A(x),A_{\nu}^B(y)]=[\Pi^{\mu}_A(x),\Pi^{\nu}_B(y)]=0\;.
\end{equation}
A state vector $|\Psi>$ is introduced that satisfies the Schrodinger equation
\begin{equation}
i\frac{d}{dt}|\Psi>=H|\Psi>
\end{equation}
where
\begin{equation}
H=\int d^dx\left[\frac{\Pi^k_A\Pi^A_k}{2\sqrt{h}}+\frac{1}{4}\sqrt{h}F^{ij}_AF^A_{ij}\right]
\label{Ham}
\end{equation}
and the constraints are imposed on the state vector
\begin{equation}
\phi_A|\Psi>=0\;\;\;\;\;\;\;\; and \;\;\;\;\;\;\;\; \chi_A|\Psi>=0\;.
\end{equation}
These conditions can also be written as $\phi_A\approx 0$ and $\chi_A\approx 0$. The constraints satisfy
\begin{equation}
[\phi_A(x),\phi_B(y)]=[\phi_A(x),\chi_B(y)]=0
\end{equation}
and
\begin{equation}
[\chi_A(x),\chi_B(y)]=igC^C_{\;\;AB}\chi_C(x)\delta(x,y)\;.
\end{equation}
This shows that these commutators vanish or weakly vanish, which is necessary for consistency.

In this approach, a gauge condition has not been imposed, but constraints follow directly from the formalism.  The constraint $\Pi^t|\Psi>=0$ gives
\begin{equation}
\frac{\delta}{\delta A_t}|\Psi>=0\;,
\end{equation}
which implies that the wave functional is independent of $A_t$. The second constraint $D_k\Pi^k_A\|\Psi>=0$ implies that the wave functional is 
invariant under time-independent gauge transformations.
Physically observable quantities must weakly commute with the constraints implying that they can depend on $A_t^A$ only through
 functions of $A_t^A$ that involve the constraints. If we neglect such operators, physically observable quantities will be independent of $A_t^A$. 
 Dynamical variables that depend on $\Pi^t_A$ will annihilate the state vector so we can restrict ourselves to dynamical variables that do not 
 depend on $A_t^A$ and $\Pi^t_A$.
 
 This theory is identical to the theory obtained by quantizing in the temporal gauge where one sets $A_t^A=0$ and
 imposes the constraint $\chi_A|\Psi>=0$ (see \cite{Hat1} and \cite{Jac1} for a discussion of the quantization of Yang-Mills theory in the temporal
 gauge in flat spacetime).
 One can, in fact, impose the constraint $A_t^A=0$ as an additional constraint in Dirac's approach. In this case, we have the first-class constraints 
 $\chi_A\approx 0$ and two sets of second class constraints $\Pi_A^t\approx 0$ and $A_t^A\approx 0$. The first class constraints are imposed on the state vector,
 as before. The second class constraints become operator constraints, $A^A_t=0$ and $\Pi_A^t= 0$, and the commutator is replaced by the Dirac bracket $[\;\; , \;\;]_D$
 where
\begin{equation}
[A_{k}^A(x),\Pi^{l}_B(y)]_D=[A_{k}^A(x),\Pi^{l}_B(y)]=i\delta^{l}_{k}\delta^A_B\delta^3(x,y)\;,
\end{equation}
\begin{equation}
[A_{k}^A(x),A_{l}^B(y)]_D=[A_{k}^A(x),A_{l}^B(y)]=0\;,
\end{equation}
and
\begin{equation}
[\Pi^{k}_A(x),\Pi^{l}_B(y)]_D=[\Pi^{k}_A(x),\Pi^{l}_B(y)]=0\;.
\end{equation}
Therefore, imposing the gauge constraint $A_t^A=0$ does not effect the theory. Thus, the Yang-Mills theory is consistently quantized in the temporal gauge.   

To complete the theory an inner product needs to be defined on the state space. If the inner product is defined as an integral over all of the $A_k^A$
it will be divergent \cite{Jac1}. Instead one defines it as an integral over physical degrees of freedom \cite{Jac1,Hen1}.
It is also possible to use the refined algebraic quantization to define 
the inner product in this system \cite{g1, g12, g21,  g2}. This can be done by first representing  all constraints in this system by   
$\check{\Lambda}_a$ ($\check{\Lambda}_a^+    =     \check{\Lambda}_a$), such that they 
satisfy  
$
[\check{\Lambda}_a; \check{\Lambda}_b] = i f^c_{ab} \check{\Lambda}_c $, 
for some structure constants $f^c_{ab}$. Now 
$L_a$, $a=\overline{1,M}$ are the  generators of the Lie  algebra, such that 
$[L_a,L_b]=if_{ab}^cL_c$. So, it is possible to define    $\mu^aL_a
\to  \exp(i\mu^aL_a)$, for the 
corresponding  Lie group  $G$. As   $\check{\Lambda}_a$  form  a
representation of      the      Lie       algebra,  
$\exp(i\mu^a\check{\Lambda}_a)$  will  form  a representation of group
$\check{T}(\exp(i\mu^aL_a)) = \exp(i\mu^a\check{\Lambda}_a)$.  The adjoint representation of the Lie algebra
can be defined as  $Ad(L_a)$, and so   $(Ad
(L_a) \rho)^c = if^c_{ab}  \rho^b$,  while  $Ad\{g\}$  is   the   adjoint
representation  of the group $(Ad\{g\} \rho)^c = (\exp(A))^c_b \rho^b$
with $A^c_b = - \mu^a f^c_{ab}$, $g = \exp(i\mu^aL_a)$.
The inner product  can now be   expressed using   the integral over gauge group \cite{g1, g12, g21,  g2}
\begin{eqnarray}
\int d_Rg (det Ad \{g\})^{-1/2} (\Phi, \check{T}(g) \Phi), 
\end{eqnarray}
where  $d_Rg$ is the right-invariant Haar measure on the group. This has been done using the 
 Giulini-Marolf group averaging formula.  Thus, it is possible to use refined algebraic quantization 
 for defining an inner product in this system.

\section{Quantization in the Lorentz Gauge}
In this section, we will demonstrate that the constraints obtained in the Lorentz gauge are the same as the temporal gauge. This is important as the ghost propagators have been studied in Lorentz gauge, and the consistency of the Yang-Mills theory in de Sitter spacetime has been demonstrated in the temporal gauge. Thus, if the quantization of Yang-Mills theory in de Sitter spacetime is not equivalent in both these gauges, it would be possible that the theory is still inconsistent in the Lorentz gauge, as it could be a gauge-related problem. It has already been argued that the quantization of Yang-Mills theory might not be well defined in certain gauges \cite{gau1ge}.  
 Now in the previous section, it was shown how the temporal gauge can be imposed as an additional constraint in Dirac's approach with the Lagrangian given by (\ref{Lag}). 
 This cannot be done with the Lorentz gauge because it contains $\dot{A}_t^A$, which cannot be written in terms of the canonical momenta (\ref{Mom}). 
As the theory is quantized in the temporal gauge, it is consistent to quantize this theory.
So, to demonstrate that it is consistent in the Lorentz gauge, we will demonstrate that the constraints obtained in the Lorentz gauge are 
  the same as the constraints obtained in the temporal gauge. 
  
  Now we can again choose  the   Lorentz gauge as the gauge fixing condition 
\begin{equation}
\nabla^{\mu}A_{\mu}^A=0\;.
\end{equation}
 To quantize in 
 the Lorentz gauge we will consider starting with the gauge-fixed Lagrangian
\begin{equation}
L_{GF}=-\frac{1}{4}\sqrt{h}\;Tr\left(F^{\mu\nu}F_{\mu\nu}\right)-\frac{1}{2}\sqrt{h}\left(\nabla^{\mu}A_{\mu}^A\right)
\left(\nabla^{\nu}A_{A\nu}\right)\;.
\end{equation}
It may be noted that this term can be related to the term used in the Feynman path integrals by integrating away the auxiliary field and also choosing 
a suitable value of $\alpha$. 
For the theory that follows from this Lagrangian to be equivalent to Yang-Mills theory the constraint $\nabla_{\mu}A^{\mu}_A\approx 0$ must be imposed. 
The canonical momenta are given by
\begin{equation}
\Pi^{\mu}_A=\sqrt{h}\left[F_A^{\mu 0}-g^{\mu t}\nabla^{\nu}A_{A\nu}\right]\;.
\end{equation}
Now
\begin{equation}
\Pi^t_A=\sqrt{h}\nabla^{\nu}A_{A\nu}\;,
\end{equation}
so there is a set of primary constraints $\phi_A$ given by
\begin{equation}
\phi_A=\Pi^t_A\approx 0\;.
\end{equation}
The canonical Hamiltonian is given by
\begin{eqnarray}
H_c &=&\int d^dx\left\{\frac{\Pi^k_A\Pi^A_k}{2\sqrt{h}}+\frac{1}{4}\sqrt{h}F^{ij}_AF^A_{ij} 
-A_0^A\chi_A  \right. \nonumber \\ && \left.  +\left[\frac{\Pi_t^A}{2\sqrt{h}}
+^{(3)}\nabla^kA_k^A-h^{ij}\dot{h}_{ij}A^A_t\right]\Pi^t_A\right\}\;.
\label{Ham2}
\end{eqnarray}
where $\chi_A=\partial_k\Pi_A^k-gC^B_{\;\; CA}A^C_k\Pi^k_B$ and
$^{(3)}\nabla$ is the covariant derivative on the three-dimensional surfaces defined by $t=$ constant. Requiring that
$\dot{\phi_A}=\{\phi_A,H_T\}\approx 0$ gives the secondary constraint
\begin{equation}
\chi_A\approx 0\;.
\end{equation}
Thus, the constraints obtained by imposing the Lorentz gauge are the same as those obtained in the previous section.

Now consider quantizing the theory: The constraints $\phi_A$ and $\chi_A$ annihilate the state vector and the state vector satisfies the Schrodinger
equation with the Hamiltonian given by (\ref{Ham}). There is an ordering ambiguity in the last term in the Hamiltonian (\ref{Ham2}), since it involves
$A_t^A$ and $\Pi^t_A$, which do not commute. We have taken the ordering as given in (\ref{Ham2}) so that $\Pi^t_A$ appears on the right and annihilates 
the state vector. Thus, quantizing the theory in the Lorentz gauge discussed here is equivalent to quantization of the theory discussed in the previous section.

\section{Conclusion}
In this paper, we have analyzed the quantization of  Yang-Mills theory in de Sitter spacetime. 
We first generalized the previous work \cite{Fai1} on the effective ghost propagators for Yang-Mills theory in de Sitter spacetime to $d+1$ dimensions. Then 
we analyzed Yang-Mills theory using Dirac constraint quantization.   
We analyzed the theory as a system of first-class constraints. We first quantized it in the temporal gauge. Then it is demonstrated that this 
analysis is consistent with quantizing this theory in the Lorentz gauge. As the constraint quantization is physically equivalent to the BRST quantization \cite{g1},  the previous work done on the BRST quantization of Yang-Mills theory in de Sitter spacetime \cite{Gibb1} is consistent. 

It would be interesting to use the results obtained in this paper to analyze different physical systems. It has been demonstrated that 
constraint quantization and the calculations done using the Feynman diagrams produce similar results \cite{feyn}. 
This was done by using a systematic expansion of all constraint equations in canonical quantum gravity. 
In fact, it was demonstrated that this method generates the conventional Feynman diagrammatic technique for graviton loops.  
It would be interesting to use this correspondence and analyze the Yang-Mills theory for different interesting physical processes. 
It was observed that the constraints obtained by imposing the Lorentz gauge are the same as those obtained when no gauge condition is imposed.
This result was expected because the theory quantized without imposing a gauge is equivalent to the theory quantized 
in the temporal gauge. However, if the theory can be consistently quantized in any gauge, then it can be transformed into a 
different gauge using the gaugeon formalism \cite{gaugeon, gaugeon1}. 
As the theory was demonstrated to be consistent in the temporal gauge, it could be converted into the Lorentz gauge by using the 
quantum gauge transformations in the gaugeon formalism. Thus, it would be interesting to analyze this system using the gaugeon formalism.

We could also use the argument used in 
  \cite{Gibb1}  and define the BRST transformations for this theory. 
  In fact, the BRST transformations can also be developed in Hamiltonian formalism using the BFV approach \cite{bfv, bfv1}. 
  Using these BRST transformations, it is possible to define finite field BRST transformations, which are a symmetry 
  of the action, but not a symmetry of the generating functional of the theory \cite{ff, ff0}. Hence, it would be interesting to 
  use the FFBRST transformations to go from the theory in the temporal gauge to the theory in the Lorentz gauge. 
  It may be noted that the relation between the first and second 
  class constraints has also been obtained using the FFBRST transformations \cite{ff1}.   
It may also be noted that the results of this paper can be used to study other interesting gauge theories on de Sitter spacetime, and 
more general geometries which have the topology $\Sigma \times R$. It would be interesting to analyze the quantization of  Chern-Simons-matter theories 
using this method. It may be noted that Chern-Simons-matter theories have been studied using the BRST transformations in different gauges \cite{ffgg, ffgg1}. 
It would be interesting to quantize the Chern-Simons-matter theory without imposing a gauge. It is expected that this will again be equivalent to
quantizing the theory in the temporal gauge. It would be interesting to analyze the equivalence of the Lorentz gauge and the temporal gauge for the quantization of Chern-Simons-matter theory. 

\section*{Acknowledgments}
M.F. would like to thank D. N. Vollick for useful discussions.

\end{document}